\documentclass[12pt]{article}
\usepackage{lscape}
\usepackage{citesort}
\usepackage{amssymb}
\usepackage{appendix}
\usepackage{multirow}

\begin{document}
\def\qq{\langle \bar q q \rangle}
\def\uu{\langle \bar u u \rangle}
\def\dd{\langle \bar d d \rangle}
\def\sp{\langle \bar s s \rangle}
\def\GG{\langle g_s^2 G^2 \rangle}
\def\Tr{\mbox{Tr}}
\def\figt#1#2#3{
        \begin{figure}
        $\left. \right.$
        \vspace*{-2cm}
        \begin{center}
        \includegraphics[width=10cm]{#1}
        \end{center}
        \vspace*{-0.2cm}
        \caption{#3}
        \label{#2}
        \end{figure}
	}
	
\def\figb#1#2#3{
        \begin{figure}
        $\left. \right.$
        \vspace*{-1cm}
        \begin{center}
        \includegraphics[width=10cm]{#1}
        \end{center}
        \vspace*{-0.2cm}
        \caption{#3}
        \label{#2}
        \end{figure}
                }

\def\ds{\displaystyle}
\def\beq{\begin{equation}}
\def\eeq{\end{equation}}
\def\bea{\begin{eqnarray}}
\def\eea{\end{eqnarray}}
\def\beeq{\begin{eqnarray}}
\def\eeeq{\end{eqnarray}}
\def\ve{\vert}
\def\vel{\left|}
\def\ver{\right|}
\def\nnb{\nonumber}
\def\ga{\left(}
\def\dr{\right)}
\def\aga{\left\{}
\def\adr{\right\}}
\def\lla{\left<}
\def\rra{\right>}
\def\rar{\rightarrow}
\def\lrar{\leftrightarrow}  
\def\nnb{\nonumber}
\def\la{\langle}
\def\ra{\rangle}
\def\ba{\begin{array}}
\def\ea{\end{array}}
\def\tr{\mbox{Tr}}
\def\ssp{{\Sigma^{*+}}}
\def\sso{{\Sigma^{*0}}}
\def\ssm{{\Sigma^{*-}}}
\def\xis0{{\Xi^{*0}}}
\def\xism{{\Xi^{*-}}}
\def\qs{\la \bar s s \ra}
\def\qu{\la \bar u u \ra}
\def\qd{\la \bar d d \ra}
\def\qq{\la \bar q q \ra}
\def\gGgG{\la g^2 G^2 \ra}
\def\GG{\langle g_s^2 G^2 \rangle}
\def\g5{\gamma_5 \not\!q}
\def\x{\gamma_5 \not\!x}
\def\g5{\gamma_5}
\def\sb{S_Q^{cf}}
\def\sd{S_d^{be}}
\def\su{S_u^{ad}}
\def\sbp{{S}_Q^{'cf}}
\def\sdp{{S}_d^{'be}}
\def\sup{{S}_u^{'ad}}
\def\ssp{{S}_s^{'??}}

\def\sig{\sigma_{\mu \nu} \gamma_5 p^\mu q^\nu}
\def\fo{f_0(\frac{s_0}{M^2})}
\def\ffi{f_1(\frac{s_0}{M^2})}
\def\fii{f_2(\frac{s_0}{M^2})}
\def\O{{\cal O}}
\def\sl{{\Sigma^0 \Lambda}}
\def\es{\!\!\! &=& \!\!\!}
\def\ap{\!\!\! &\approx& \!\!\!}
\def\ar{&+& \!\!\!}
\def\arrr{\!\!\!\! &+& \!\!\!}
\def\ek{&-& \!\!\!}
\def\vev{&\vert& \!\!\!}
\def\kek{\!\!\!\!&-& \!\!\!}
\def\cp{&\times& \!\!\!}
\def\se{\!\!\! &\simeq& \!\!\!}
\def\eqv{&\equiv& \!\!\!}
\def\kpm{&\pm& \!\!\!}
\def\kmp{&\mp& \!\!\!}
\def\mcdot{\!\cdot\!}
\def\erar{&\rightarrow&}


\def\simlt{\stackrel{<}{{}_\sim}}
\def\simgt{\stackrel{>}{{}_\sim}}


\title{
         {\Large
                 {\bf
Transition magnetic moments between negative parity heavy baryons
                 }
         }
      }

\author{\vspace{1cm}\\
{\small T. M. Aliev \thanks {e-mail:
taliev@metu.edu.tr}~\footnote{permanent address:Institute of
Physics,Baku,Azerbaijan}\,\,, T. Barakat \thanks {e-mail:
tbarakat@KSU.EDU.SA}\,\,, M. Savc{\i} \thanks
{e-mail: savci@metu.edu.tr}} \\
{\small Physics Department, Middle East Technical University,
06531 Ankara, Turkey }\\
{\small $^\ddag$ Physics and Astronomy Department, King Saud University, Saudi Arabia}}

\date{}

\begin{titlepage}
\maketitle
\thispagestyle{empty}

\begin{abstract}

The transition magnetic moments between negative parity, spin-1/2 heavy
baryons are studied in framework of the light cone QCD sum rules.
By constructing the sum rules for different Lorentz structures, the unwanted
contributions coming from negative (positive) to positive (negative) parity
transitions are removed. It is found that the magnetic moments between
neutral negative parity heavy $\Xi_Q^{\prime 0}$ and $\Xi_Q^0$ baryons are
very small. Magnetic moments of the $\Sigma_Q \to \Lambda_Q$ and $
\Xi_Q^{\prime \pm} \to \Xi_Q^\pm$ transitions are quite large and can be
measured in further experiments. 

\end{abstract}

~~~PACS numbers: 11.55.Hx, 13.40.Em, 14.20.Lq, 14.20.Mr,

\end{titlepage}

\section{Introduction}

Study of the electromagnetic properties of baryons constitutes very important 
source of information in exploring their internal structure, and can
provide valuable insight in understanding the nonperturbative aspect of QCD.
In recent years, study of the properties of heavy baryons has become a
subject of growing interest due to the experimental observation of many
heavy baryons. All ground state baryons containing single charm and bottom
quarks, except $\Omega_b^\ast$ baryon, are observed and their masses are
measured (for a review, see \cite{Rfrd01}). Moreover, a number of negative
parity baryons have also been observed.

These exciting experimental results have stimulated the theoretical studies
along these lines. The mass and magnetic moments of the heavy baryons can
serve useful information about their inner structure. Experimentally the
magnetic moments of all members of the octet $J^P={1^+\over 2}$ baryons
(except $\Sigma^0$ baryon), and two members of the decuplet
$J^P={3^+\over 2}$ baryons are measured \cite{Rfrd01}.

The magnetic moments of the $J^P={1^+\over 2}$ light and heavy baryons have
extensively been calculated in numerous theoretical approaches. The
approaches based on naive quark model in \cite{Rfrd02,Rfrd03},
relativistic quark model\cite{Rfrd04}, nonrelativistic quark
model\cite{Rfrd05}, chiral quark model \cite{Rfrd06},
chiral perturbation theory \cite{Rfrd07}, hypercentral model \cite{Rfrd08},
soliton model \cite{Rfrd09}
traditional \cite{Rfrd10} and light version of the QCD sum rules method
(LCSR) \cite{Rfrd11,Rfrd12,Rfrd13,Rfrd14} have
been employed in studying the masses and magnetic
moments of the heavy baryons. The magnetic moments of the negative parity
heavy baryons has recently been considered within the framework of the LCSR
method in \cite{Rfrd15}.

In this work, we extend our analysis to determine the magnetic
moments for the $\Sigma_Q \to \Lambda_Q$ and $\Xi_Q^\prime \to \Xi_Q$
transitions between the negative parity baryons.

In the following section we derive the light cone sum rules for the
magnetic moments of the aforementioned transitions. Section 3 is devoted to
the numerical analysis of the obtained sum rules for the transition
magnetic moments. A comparison of our predictions with the results from
other approaches is given also.

\section{Theoretical framework}
Following the philosophy of the QCD sum rules,
the magnetic moments of the $\Sigma_Q \to \Lambda_Q$ and $\Xi_Q^\prime \to
\Xi_Q$ transitions of the baryons with negative parity can be obtained in
LCSR by matching two representations of the relevant correlation function
written in terms of the hadronic and quark-gluon languages. For this purpose
we use the correlation function
\bea
\label{efrd01}
\Pi (p,q) = i \int d^4x e^{ipx} \lla 0 \vel \mbox{\rm T} \{
\eta_Q (x) \bar{\eta}_Q(0) \}\ver 0 \rra_\gamma ~,
\eea
where $\gamma$ is the external electromagnetic field, $\eta_Q$ is the
interpolating current of the heavy baryon with single heavy quark.
This correlation function can be calculated at hadronic level by inserting a
complete set of hadrons carrying the same quantum numbers of the correlation
function, and isolating the contributions arising from the ground states
which have poles in $p^2$ and $(p+q)^2$. The interpolating current can
interact with both negative and positive parity baryons, therefore it can be
written in the following form,
\bea
\label{efrd02}
\Pi(p,q) \es
{\la 0 \ve \eta \ve B_2^{(+)}(p,s) \ra \over p^2-m_{B_2^{(+)}}^2} \la B_2^{(+)}(p,s)
\gamma(q)\ve
B_1^{(+)}(p+q,s) \ra {\la B_1^{(+)}(p+q,s) \ve \bar{\eta}(0) \ra \over
(p+q)^2-m_{B_1^{(+)}}^2} \nnb \\
\ar {\la 0 \ve \eta \ve B_2^{(-)}(p,s) \ra \over p^2-m_{B_2^{(-)}}^2} \la B_2^{(-)}(p,s)
\gamma(q) \ve
B_1^{(-)}(p+q,s) \ra {\la B_1^{(-)}(p+q,s) \ve \bar{\eta}(0) \ra \over
(p+q)^2-m_{B_1^{(-)}}^2} \nnb \\
\ar {\la 0 \ve \eta \ve B_2^{(+)}(p,s) \ra \over p^2-m_{B_2^{(+)}}^2} \la B_2^{(+)}(p,s)
\gamma(q)\ve
B_1^{(-)}(p+q,s) \ra {\la B_1^{(-)}(p+q,s) \ve \bar{\eta}(0) \ra \over
(p+q)^2-m_{B_1^{(-)}}^2} \nnb \\
\ar {\la 0 \ve \eta \ve B_2^{(-)}(p,s) \ra \over p^2-m_{B_2^{(-)}}^2} \la B_2^{(-)}(p,s)
\gamma(q)\ve
B_1^{(+)}(p+q,s) \ra {\la B_1^{(+)}(p+q,s) \ve \bar{\eta}(0) \ra \over
(p+q)^2-m_{B_1^{(+)}}^2} + \cdots~,
\eea
where $B^{(\pm)}$ and $m_{B^{(\pm)}}$ correspond to positive (negative)
parity baryons and their masses, respectively; $q$ is the photon momentum;
and dots correspond to the higher states contributions.

The matrix elements in Eq. (\ref{efrd02}) are defined as,
\bea
\label{efrd03}
\la 0 \ve \eta \ve B^{(+)}(p)\ra \es  \lambda_{B^{(+)}} u^{(+)} (p)~,\nnb \\
\la 0 \ve \eta \ve B^{(-)}(p)\ra \es  \lambda_{B^{(-)}} \gamma_5 u^{(-)} (p)~,\nnb \\
\la B_2^{(+)}(p) \gamma(q) \ve \eta \ve B_1^{(+)}(p+q)\ra \es e \varepsilon^\mu
\bar{u}^{(+)}(p)
\left[\gamma_\mu f_1 - {i \sigma_{\mu\nu} q^\nu \over  m_{B_1^{(+)}}
+ m_{B_2^{(+)}} } f_2 \right] u^{(+)}(p+q) \nnb \\
\es e \varepsilon^\mu \bar{u}^{(+)}(p) \left[ (f_1+f_2) \gamma_\mu - {(2p+q)_\mu
\over m_{B_1^{(+)}} + m_{B_2^{(+)}} } f_2 \right] u^{(+)}(p+q)~, \nnb \\
\la B_2^{(-)}(p) \gamma(q) \ve \eta \ve B_1^{(+)}(p+q)\ra \es e \varepsilon^\mu
\bar{u}^{(-)}(p)
\left[\gamma_\mu f_1^T - {i \sigma_{\mu\nu} q^\nu \over 
m_{B_1^{(+)}} + m_{B_2^{(-)}} } f_2^T
\right] \gamma_5 u^{(+)}(p+q) \nnb \\
\es e \varepsilon^\mu \bar{u}^{(-)}(p)
\left[\left( f_1^T - {m_{B_1^{(+)}} - m_{B_2^{(-)}} \over 
m_{B_1^{(+)}} + m_{B_2^{(-)}} } f_2^T \right) \gamma_\mu \right. \nnb \\
\ek \left. {(2p+q)_\mu \over m_{B_1^{(+)}} + m_{B_2^{(-)}} } f_2^T \right] \gamma_5
u^{(+)}(p+q)~, \nnb \\ 
\la B_2^{(-)}(p) \gamma(q) \ve \eta \ve B_1^{(-)}(p+q)\ra \es e \varepsilon^\mu
\bar{u}^{(-)}(p)
\left[\gamma_\mu f_1^\ast - {i \sigma_{\mu\nu} q^\nu \over 
m_{B_1^{(-)}} + m_{B_2^{(-)}} } f_2^\ast
\right] u^{(-)}(p+q) \nnb \\
\es e \varepsilon^\mu \bar{u}^{(-)}(p) \left[\left(f_1^\ast + f_2^\ast
\right) \gamma_\mu - {(2p+q)_\mu \over m_{B_1^{(-)}} + m_{B_2^{(-)}} }
f_2^\ast \right] u^{(-)}(p+q)~,
\eea    
where $\varepsilon^\mu$ is the four-polarization vector.

Performing summation over spins of the
heavy baryons, for the correlation function from the hadronic side we get,
\bea
\label{efrd04} 
\Pi (p,q) \es A ({\not\!{p}}_2 +  m_{B_2^{(+)}}) 
\not\!{\varepsilon} ({\not\!{p}}_1 +  m_{B_1^{(+)}})~\nnb \\
\ar B ({\not\!{p}}_2 -  m_{B_2^{(-)}})       
\not\!{\varepsilon} ({\not\!{p}}_1 -  m_{B_1^{(-)}})~\nnb \\
\ar C  ({\not\!{p}}_2 -  m_{B_2^{(-)}})       
\not\!{\varepsilon} \gamma_5({\not\!{p}}_1 +  m_{B_1^{(+)}})~\nnb\\
\ar D  ({\not\!{p}}_2 +  m_{B_2^{(+)}})       
\not\!{\varepsilon} \gamma_5({\not\!{p}}_1 -  m_{B_1^{(-)}})~,
\eea
where
\bea
\label{efrd05}
A\es { \lambda_{B_1^{(+)}}  \lambda_{B_2^{(+)}} (f_1+f_2) \over
( m_{B_1^{(+)}}^2-p_1^2)( m_{B_2^{(+)}}^2-p_2^2)} \nnb \\
B\es { \lambda_{B_1^{(-)}}  \lambda_{B_2^{(-)}} (f_1^\ast+f_2^\ast) \over
( m_{B_1^{(-)}}^2-p_1^2)( m_{B_2^{(-)}}^2-p_2^2)} \nnb \\
C\es { \lambda_{B_1^{(-)}} \lambda_{B_2^{(+)}} \over
( m_{B_1^{(-)}}^2-p_1^2)( m_{B_2^{(+)}}^2-p_2^2)}
\left[f_1^T + { m_{B_1^{(-)}} -  m_{B_2^{(+)}} \over m_{B_1^{(-)}} + m_{B_2^{(+)}}} f_2^T \right]\nnb \\
D\es { \lambda_{B_1^{(+)}} \lambda_{B_2^{(-)}} \over
( m_{B_1^{(+)}}^2-p_1^2)( m_{B_2^{(-)}}^2-p_2^2)}
\left[f_1^T - { m_{B_1^{(+)}} - m_{B_2^{(-)}} \over   m_{B_1^{(+)}} + m_{B_2^{(-)}} } f_2^T \right]~,
\eea
where $p_1 = p+q$ and $p_2=p$.

Among the terms in Eq. (\ref{efrd04})
\bea
\label{nolabel01}
f_1+f_2~,~~(f_1^\ast+f_2^\ast)~,~~f_1^T + { m_{B_1^{(-)}} - m_{B_2^{(+)}}
\over  m_{B_1^{(-)}} + m_{B_2^{(+)}} } f_2^T~,~~f_1^T - { m_{B_1^{(+)}} -
m_{B_2^{(-)}} \over m_{B_1^{(+)}} + m_{B_2^{(-)}} } f_2^T~,\nnb
\eea
that are proportional to $\gamma_\mu$, the first two correspond
to the magnetic moments of the positive to
positive, negative to negative transitions, respectively;
and the third and the fourth ones describe the transition magnetic moments between positive and
negative parity baryons at $q^2=0$.
Our aim in the present work
is to calculate the transition magnetic moment between the negative parity
baryons, and therefore we should find a way to remove the other three contributions.

In order to determine the transition magnetic moments between negative
parity baryons four equations are needed, for which we choose the following
four Lorentz structures, $(\varepsilon\!\cdot\! p) I$,
$(\varepsilon\!\cdot\! p) \rlap/{p}$, $\rlap/{p}\rlap/{\varepsilon}$ and
$\rlap/{\varepsilon}$. Solving finally these four coupled equations, we
obtain the unknown coefficient $B$ which describes the negative to negative
parity transition.

It follows from Eq. (\ref{efrd01}) that interpolating currents are needed
in order to calculate the correlation function in terms of quarks and
gluons. Here, it should be remembered that hadrons containing single heavy
quark belong to either sextet or anti-triplet representations of $SU(3)$.
Sextet (anti-triplet) representation is symmetric (antisymmetric) with
respect to the exchange of light quarks. In constructing the interpolating
currents belonging to sextet and anti-triplet representations we will use
this fact, whose explicit forms are given as (see \cite{Rfrd16}),
\bea
\label{efrd06}
\eta^{(s)} \es -{1\over \sqrt{2}} \varepsilon^{abc} \Big\{
(q_1^{aT} C Q^b) \gamma_5 q_2^c +
t (q_1^{aT} C \gamma_5 Q^b) q_2^c +
(q_2^{aT} C Q^b) \gamma_5 q_1^c + (q_2^{aT} C
\gamma_5 Q^b) q_1^c\Big\}~, \nnb \\
\eta^{(a)} \es -{1\over \sqrt{6}} \varepsilon^{abc} \Big\{
2 (q_1^{aT} C q_2^b) \gamma_5 Q^c +
2 t (q_1^{aT} C \gamma_5 q_2^b) Q^c +
(q_1^{aT} C Q^b) \gamma_5 q_2^c \nnb \\
\ar t (q_1^{aT} C \gamma_5 Q^b) q_2^c -
(q_2^{aT} C Q^b) \gamma_5 q_1^c -
t (q_2^{aT} C \gamma_5 Q^b) q_1^c\Big\}~.
\eea
In this expression $a,b,c$ are the color indices; $C$ is the
charge conjugation operator; and $t$ is a free parameter (the choice $t=-1$
correspond to the Ioffe current). The light quark contents of the heavy
$\Sigma_Q$, $\Xi_Q^\prime$, $\Lambda_Q$ and $\Xi_Q$ baryons are presented in
Table (1).


\begin{table}[h]

\renewcommand{\arraystretch}{1.3}
\addtolength{\arraycolsep}{-0.5pt}
\small
$$
\begin{array}{|c|c|c|c|c|c|c|c|c|}
\hline \hline   
 & \Sigma_{c(b)}^{+(0)} & \Sigma_{c(b)}^{0(-)} &
             \Xi_{c(b)}^{\prime 0(-)}  & \Xi_{c(b)}^{\prime +(0)}  &
 \Lambda_{c(b)}^{+(0)} & 
\Xi_{c(b)}^{0(-)}  & \Xi_{c(b)}^{+(0)} \\   
\hline \hline
q_1   & u & d & d & u & u & d & u \\
q_2   & d & d & s & s & d & s & s \\
\hline \hline 
\end{array}
$$
\caption{Light quark contents of the heavy
$\Sigma_Q$, $\Xi_Q^\prime$, $\Lambda_Q$ and $\Xi_Q$ baryons.}
\renewcommand{\arraystretch}{1}
\addtolength{\arraycolsep}{-1.0pt}
\end{table}      


Having the interpolating currents containing single heavy baryon 
at hand, the correlation function can easily be calculated. The correlation
functions describing the sextet to anti-triplet transition magnetic moments
in the light cone version of the sum rules contain the following
contributions. The photon interacts with light or heavy quarks
perturbatively. This contribution can be obtained by replacing one of the
free quark operators by,

\bea
\label{efrd07}
S^{free} (x) \to -{1\over 2}\int d^4y S^{free} (x-y) \gamma^\mu S^{free} (y)
y^\nu {\cal F}_{\mu\nu}~,
\eea
and the other two propagators are taken as the free quark operator. In Eq.
(\ref{efrd07}) the Fock-Schwinger gauge, i.e., $A_\mu = {1\over 2}
{\cal F}_{\mu\nu} y^\nu$ has been used.
The other type of contribution can be calculated by replacing one of the
propagators in the same manner as is given in Eq. (\ref{efrd07}), and replacing
the other one (or both) by the ``full" light quark operator.

Last type of contribution is the nonperturbative one, that can be obtained
by replacing one of the light quark operators with,
\bea                                                                       
\label{efrd08}
S_{\mu\nu}^{ab} \to -{1\over 4} \left( q^a \Gamma_i q^b \right)
\left(\Gamma_i \right)_{\mu\nu},
\eea 
where $\Gamma_i$ are the full set of Dirac matrices; and the remaining two
other propagators are taken as the full quark propagators. In calculating
these contributions, the expressions of the light and heavy quark
propagators in external field are needed. The light cone expansion of the
propagator in external field is performed in \cite{Rfrd17}, and it is found
that the contributions of the three-particle $\bar{q}G q$, and the four-particle
$\bar{q}G^2 q$, $\bar{q} q \bar{q}q$ nonlocal operators are small. Keeping
this fact in mind, the expressions of the light and heavy quark propagators
in external field are given by,
\bea
\label{efrd09}
S_q(x) \es {i \rlap/x\over 2\pi^2 x^4} - {m_q\over 4 \pi^2 x^2} -
{\lla \bar q q \rra\over 12} \left(1 - i {m_q\over 4} \rlap/x \right) -
{x^2\over 192} m_0^2 \lla \bar q q \rra  \left( 1 -
i {m_q\over 6}\rlap/x \right) \nnb \\
\ek i g_s \int_0^1 du \Bigg[{\rlap/x\over 16 \pi^2 x^2} G_{\mu \nu} (ux)
\sigma_{\mu \nu} - {i\over 4 \pi^2 x^2} u x^\mu G_{\mu \nu} (ux) \gamma^\nu \nnb \\
\ek i {m_q\over 32 \pi^2} G_{\mu \nu} \sigma^{\mu
 \nu} \left( \ln {-x^2 \Lambda^2\over 4}  +
 2 \gamma_E \right) \Bigg]~, \nnb \\ \nnb \\
S_Q(x) \es {m_Q^2 \over 4 \pi^2} \Bigg\{ {K_1(m_Q\sqrt{-x^2}) \over \sqrt{-x^2}} +
i {\rlap/{x} \over (\sqrt{-x^2})^2} K_2(m_Q\sqrt{-x^2}) \Bigg\} \nnb \\ 
\ek {g_s \over 16 \pi^2} \int_0^1 du
G_{\mu\nu}(ux) \left[ \left(\sigma^{\mu\nu} \rlap/x + \rlap/x
\sigma^{\mu\nu}\right) {K_1 (m_Q\sqrt{-x^2})\over \sqrt{-x^2}} +
2 \sigma^{\mu\nu} K_0(m_Q\sqrt{-x^2})\right]~.
\eea
Here $\Lambda$ is the cut-off energy separating the perturbative and
nonperturbative regions, $K_i$ are the modified Bessel functions.

It should be noted here that, in the expression (\ref{efrd08}) which is used
to calculate nonperturbative contributions, there appears the matrix
elements of the nonlocal operators between vacuum and one photon states of
the form $\la \gamma(q) \ve \bar{q} \Gamma_i q \ve 0 \ra$, in which all
nonperturbative effects are encoded. These matrix elements are given in
terms of the photon distribution amplitudes as \cite{Rfrd18},

\bea
\label{efrd10}
&&\langle \gamma(q) \vert  \bar q(x) \sigma_{\mu \nu} q(0) \vert  0
\rangle  = -i e_q \bar q q (\varepsilon_\mu q_\nu - \varepsilon_\nu
q_\mu) \int_0^1 du e^{i \bar u qx} \left(\chi \varphi_\gamma(u) +
\frac{x^2}{16} \mathbb{A}  (u) \right) \nnb \\ &&
-\frac{i}{2(qx)}  e_q \qq \left[x_\nu \left(\varepsilon_\mu - q_\mu
\frac{\varepsilon x}{qx}\right) - x_\mu \left(\varepsilon_\nu -
q_\nu \frac{\varepsilon x}{q x}\right) \right] \int_0^1 du e^{i \bar
u q x} h_\gamma(u)
\nnb \\
&&\langle \gamma(q) \vert  \bar q(x) \gamma_\mu q(0) \vert 0 \rangle
= e_q f_{3 \gamma} \left(\varepsilon_\mu - q_\mu \frac{\varepsilon
x}{q x} \right) \int_0^1 du e^{i \bar u q x} \psi^v(u)
\nnb \\
&&\langle \gamma(q) \vert \bar q(x) \gamma_\mu \gamma_5 q(0) \vert 0
\rangle  = - \frac{1}{4} e_q f_{3 \gamma} \epsilon_{\mu \nu \alpha
\beta } \varepsilon^\nu q^\alpha x^\beta \int_0^1 du e^{i \bar u q
x} \psi^a(u)
\nnb \\
&&\langle \gamma(q) | \bar q(x) g_s G_{\mu \nu} (v x) q(0) \vert 0
\rangle = -i e_q \qq \left(\varepsilon_\mu q_\nu - \varepsilon_\nu
q_\mu \right) \int {\cal D}\alpha_i e^{i (\alpha_{\bar q} + v
\alpha_g) q x} {\cal S}(\alpha_i)
\nnb \\
&&\langle \gamma(q) | \bar q(x) g_s \tilde G_{\mu \nu} i \gamma_5 (v
x) q(0) \vert 0 \rangle = -i e_q \qq \left(\varepsilon_\mu q_\nu -
\varepsilon_\nu q_\mu \right) \int {\cal D}\alpha_i e^{i
(\alpha_{\bar q} + v \alpha_g) q x} \tilde {\cal S}(\alpha_i)
\nnb \\
&&\langle \gamma(q) \vert \bar q(x) g_s \tilde G_{\mu \nu}(v x)
\gamma_\alpha \gamma_5 q(0) \vert 0 \rangle = e_q f_{3 \gamma}
q_\alpha (\varepsilon_\mu q_\nu - \varepsilon_\nu q_\mu) \int {\cal
D}\alpha_i e^{i (\alpha_{\bar q} + v \alpha_g) q x} {\cal
A}(\alpha_i)
\nnb \\
&&\langle \gamma(q) \vert \bar q(x) g_s G_{\mu \nu}(v x) i
\gamma_\alpha q(0) \vert 0 \rangle = e_q f_{3 \gamma} q_\alpha
(\varepsilon_\mu q_\nu - \varepsilon_\nu q_\mu) \int {\cal
D}\alpha_i e^{i (\alpha_{\bar q} + v \alpha_g) q x} {\cal
V}(\alpha_i) \nnb \\ && \langle \gamma(q) \vert \bar q(x)
\sigma_{\alpha \beta} g_s G_{\mu \nu}(v x) q(0) \vert 0 \rangle  =
e_q \qq \left\{
        \left[\left(\varepsilon_\mu - q_\mu \frac{\varepsilon x}{q x}\right)\left(g_{\alpha \nu} -
        \frac{1}{qx} (q_\alpha x_\nu + q_\nu x_\alpha)\right) \right. \right. q_\beta
\nnb \\ && -
         \left(\varepsilon_\mu - q_\mu \frac{\varepsilon x}{q x}\right)\left(g_{\beta \nu} -
        \frac{1}{qx} (q_\beta x_\nu + q_\nu x_\beta)\right) q_\alpha
\nnb \\ && -
         \left(\varepsilon_\nu - q_\nu \frac{\varepsilon x}{q x}\right)\left(g_{\alpha \mu} -
        \frac{1}{qx} (q_\alpha x_\mu + q_\mu x_\alpha)\right) q_\beta
\nnb \\ &&+
         \left. \left(\varepsilon_\nu - q_\nu \frac{\varepsilon x}{q.x}\right)\left( g_{\beta \mu} -
        \frac{1}{qx} (q_\beta x_\mu + q_\mu x_\beta)\right) q_\alpha \right]
   \int {\cal D}\alpha_i e^{i (\alpha_{\bar q} + v \alpha_g) qx} {\cal T}_1(\alpha_i)
\nnb \\ &&+
        \left[\left(\varepsilon_\alpha - q_\alpha \frac{\varepsilon x}{qx}\right)
        \left(g_{\mu \beta} - \frac{1}{qx}(q_\mu x_\beta + q_\beta x_\mu)\right) \right. q_\nu
\nnb \\ &&-
         \left(\varepsilon_\alpha - q_\alpha \frac{\varepsilon x}{qx}\right)
        \left(g_{\nu \beta} - \frac{1}{qx}(q_\nu x_\beta + q_\beta x_\nu)\right)  q_\mu
\nnb \\ && -
         \left(\varepsilon_\beta - q_\beta \frac{\varepsilon x}{qx}\right)
        \left(g_{\mu \alpha} - \frac{1}{qx}(q_\mu x_\alpha + q_\alpha x_\mu)\right) q_\nu
\nnb \\ &&+
         \left. \left(\varepsilon_\beta - q_\beta \frac{\varepsilon x}{qx}\right)
        \left(g_{\nu \alpha} - \frac{1}{qx}(q_\nu x_\alpha + q_\alpha x_\nu) \right) q_\mu
        \right]
    \int {\cal D} \alpha_i e^{i (\alpha_{\bar q} + v \alpha_g) qx} {\cal T}_2(\alpha_i)
\nnb \\ &&+
        \frac{1}{qx} (q_\mu x_\nu - q_\nu x_\mu)
        (\varepsilon_\alpha q_\beta - \varepsilon_\beta q_\alpha)
    \int {\cal D} \alpha_i e^{i (\alpha_{\bar q} + v \alpha_g) qx} {\cal T}_3(\alpha_i)
\nnb \\ &&+
        \left. \frac{1}{qx} (q_\alpha x_\beta - q_\beta x_\alpha)
        (\varepsilon_\mu q_\nu - \varepsilon_\nu q_\mu)
    \int {\cal D} \alpha_i e^{i (\alpha_{\bar q} + v \alpha_g) qx} {\cal T}_4(\alpha_i)
                        \right\}~,
\eea
where $\varphi_\gamma(u)$ is the leading twist-2, $\psi^v(u)$,
$\psi^a(u)$, ${\cal A}$ and ${\cal V}$ are the twist-3, and
$h_\gamma(u)$, $\mathbb{A}$, ${\cal T}_i$ ($i=1,~2,~3,~4$) are the
twist-4 photon DAs, and $\chi$ is the magnetic susceptibility.
The measure ${\cal D} \alpha_i$ is defined as
\bea
\label{nolabel05}
\int {\cal D} \alpha_i = \int_0^1 d \alpha_{\bar q} \int_0^1 d
\alpha_q \int_0^1 d \alpha_g \delta(1-\alpha_{\bar
q}-\alpha_q-\alpha_g)~.\nnb
\eea

As has already been noted, in determination of the magnetic moment
responsible for the negative to negative parity transition, four equations
are needed, and for this purpose we choose the coefficients of the
structures $(\varepsilon\!\cdot\! p) I$,
$(\varepsilon\!\cdot\! p) \rlap/{p}$, $\rlap/{p}\rlap/{\varepsilon}$ and  
$\rlap/{\varepsilon}$. The sum rules for the negative to negative parity
transition magnetic moments can be obtained by choosing the coefficients of
the aforementioned Lorentz structures $\Pi_i$, and equate them to the
corresponding coefficients in hadronic part. Solving then the linear
equations for the coefficients describing the negative to negative
transition magnetic moments, and performing Borel transformation over the
variables $-p^2$ and $-(p+q)^2$ in order to suppress higher states and
continuum contribution, we finally obtain the magnetic moment for the negative
to negative parity baryon transitions as is given below,
\bea                                                                        
\label{efrd11}
\mu \es {e^{m_{B_1^{(-)}}^2/2M^2} e^{m_{B_2^{(-)}}^2/2M^2}  \over
2 \lambda_{B_1^{(-)}}  \lambda_{B_2^{(-)}}
\left( m_{B_1^{(+)}} +  m_{B_1^{(-)}}\right)
\left( m_{B_2^{(+)}} +  m_{B_2^{(-)}}\right)}
\Big\{ \left( m_{B_1^{(+)}} + m_{B_2^{(-)}}\right)
\left(\Pi_1^{B} - m_{B_2^{(+)}} \Pi_2^{B}\right) \nnb \\
\ar 2 m_{B_2^{(+)}} \Pi_3^{B} - 2 \Pi_4^{B}\Big\}~.
\eea
In this expression we take $M_1^2=M_2^2=2 M^2$, since the masses of the
$\Sigma_Q$, $\Lambda_Q$, $\Xi_Q^\prime$ and $\Xi_Q$ baryons are very close
to each other. The expressions of $\Pi_i^B$ are presented in Appendix A.

It follows from Eq. (\ref{efrd11}) that in determination of the magnetic
moments of the $\Sigma_Q \to \Lambda_Q$ and $\Xi_Q^\prime \to \Xi_Q$
transitions, the residues of the negative parity heavy baryons are
necessary. These residues can be determined from the analysis of the
two-point correlation function
\bea
\label{nolabel06} 
\Pi(q^2)= i \int d^4x e^{iqx} \la 0 \ve \mbox{\rm T} \left\{ \eta_Q(x)
\bar{\eta}_Q(0) \right\} \ve 0 \ra~, \nnb
\eea
where $\eta_Q$ is the interpolating current for the corresponding heavy
baryon given by Eq. (\ref{efrd06}). This interpolating current interacts
with both positive and negative parity heavy baryons. Saturating this
correlation function with the ground states of positive and negative parity
baryons we have,
\bea
\label{nolabel07}
\Pi (q^2) = {\ve \lambda_{B^{(-)}} \ve^2 (\not\!p - m_{B^{(-)}}) \over
m_{B^{(-)}}^2-p^2} +
{\ve \lambda_{B^{(+)}} \ve^2 (\not\!p + m_{B^{(+)}}) \over m_{B^{(+)}}^2-p^2}~\nnb .
\eea
Eliminating the contributions coming from the positive parity baryons, the
following sum rules for the residue and mass of the negative parity baryons
are obtained,
\bea
\label{nolabel08}
\ve \lambda_{B^{(-)}} \ve^2 \es {1\over \pi} {e^{m_{B^{(-)}}^2/M^2} \over m_{B^{(+)}}+m_{B^{(-)}}} 
\int ds e^{-s/M^2} \left[ m_{B^{(+)}} \mbox{\rm Im} \Pi_1^M(s) - \mbox{\rm Im}
\Pi_2^M(s) \right]~,\nnb \\
\label{nolabel09}
m_{B^{(-)}}^2 \es {\int_{m_Q^2}^{s_0} s ds e^{-s/M^2} \left[  m_{B^{(+)}} \mbox{\rm Im}
\Pi_1^M(s) - \mbox{\rm Im} \Pi_2^M(s) \right] \over
\int_{m_Q^2}^{s_0} ds e^{-s/M^2} \left[  m_{B^{(+)}} \mbox{\rm Im}     
\Pi_1^M(s) - \mbox{\rm Im} \Pi_2^M(s) \right]}~ \nnb.
\eea
Here $\Pi_1^M$ and $\Pi_2^M$ are the invariant functions corresponding to the
structures $\not\!\!p$ and $I$, respectively. The expressions of $\Pi_1^M$ and
$\Pi_2^M$ for the $\Sigma_Q^0$ baryon are presented in Appendix B.

\section{Numerical analysis}

In this section we present our numerical results for the magnetic moments of
the $\Sigma_Q \to \Lambda_Q$ and $\Xi_Q^\prime \to \Xi_Q$ transitions for
negative parity baryons derived from the LCSR. In this numerical analysis
the values of the the values of the relevant input parameters entering to
the LCSR are needed. The main nonperturbative input of LCSR is the DAs which
are all calculated in \cite{Rfrd18}, and for completeness we present
their expressions in Appendix C. The other input parameters needed in the
numerical analysis are, quark condensate $\qq$, $m_0^2$, magnetic
susceptibility $\chi$ of quarks, etc. In further numerical calculations
we use $\left[ \uu = \dd \right]_{\mu=1~GeV} = -(0.243)^3~GeV^3$
\cite{Rfrd19}, $\sp \ve_{\mu=1~GeV} = 0.8 \uu\ve_{\mu=1~GeV}$,
$m_0^2=(0.8\pm 0.2)~GeV^2$ \cite{Rfrd20}. The magnetic susceptibility was
determined within the QCD sum rules in \cite{Rfrd21,Rfrd22,Rfrd23}).

Having all necessary ingredients at hand, we are now ready to perform
the numerical analysis for the transition magnetic moments of the negative
parity baryons. Sum rules contain also three auxiliary parameters in the
interpolating current other than
those input parameters given above: Borel mass parameter $M^2$, continuum
threshold $s_0$, and the arbitrary parameter $t$. We demand that the
magnetic moment should be independent on these auxiliary parameters.
Therefore we shall look for the ``working regions" of these parameters,
where magnetic moments exhibit good stability with respect to their variations
in respective domains. It should be remembered  that the continuum  threshold
$s_0$ is not arbitrary but related to the first excited states. The
difference $\sqrt{s_0}-m_{ground}$ is the energy needed to transfer the
baryon to its first excited state. Usually this difference varies in the
range $0.3~GeV \le \sqrt{s_0}-m_{ground} \le 0.8~GeV$, and in our analysis
we choose the average value $\sqrt{s_0}-m_{ground}=0.5~GeV$.

Having determined the value of $s_0$, next we try to find the ``working
regions" of the Borel parameter $M^2$. The upper bound of $M^2$ is obtained
by demanding that contributions of higher states and continuum constitute
about 40\% of the perturbative part.The lower bound is determined from the
condition that higher twist contributions are less than the leading twist
contributions. Our analysis shows that the working regions of $M^2$ where
both conditions are satisfied are
\bea
\label{nolabel09}
2.5~GeV^2 \le M^2 \le 4.0~GeV^2,~~\mbox{for $\Sigma_c$, $\Xi_c^\prime$,
$\Lambda_c$, $\Xi_c$}~, \nnb \\
4.5~GeV^2 \le M^2 \le 7.0~GeV^2,~~\mbox{for $\Sigma_b$, $\Xi_b^\prime$,
$\Lambda_b$, $\Xi_b$}~. \nnb
\eea
As an example, in Figs. (1) and (2) we present the dependence of
$\mu_{\Sigma_b^0 \to \Lambda_b^0}$ on $M^2$ at several fixed values of the
arbitrary parameter $t$, at $s_0=40.0~GeV^2$ and  $s_0=42.5~GeV^2$. We
observe from these figures that transition magnetic moment exhibits good
stability when $M^2$ varies in the region $3.0~GeV^2 \le M^2 \le 4.0~GeV^2$.   
In Figs. (3) and (4) we present the dependence of
$\mu_{\Sigma_b^0 \to \Lambda_b^0}$ on $\cos\theta$ (where $t=\tan\theta$ at
two fixed values of $M^2$ and at $s_0=40.0~GeV^2$ and  $s_0=42.5~GeV^2$,
respectively. We see from these figures that, when $\cos\theta$ varies in
the domain $-1.0 \le \cos\theta \le -0.7$, the magnetic moment demonstrates
good stability with respect to the variation in $\cos\theta$. Our final
result for the transition magnetic moment is $\mu_{\Sigma_b^0 \to
\Lambda_b^0}=(-0.3 \pm 0.05) \mu_N$.

The analysis of the sum rules for the other transition magnetic moments are
also calculated, whose values can be summarized as,
\bea
\label{nolabel10}
\mu_{\Sigma_c^+ \to \Lambda_c^+} \es (0.25 \pm 0.05) \mu_N~, \nnb \\
\mu_{\Xi_c^{\prime 0} \to \Xi_c^{0}} \es (0.08 \pm 0.01) \mu_N~, \nnb \\
\mu_{\Xi_c^{\prime +} \to \Xi_c^{+}} \es (0.20 \pm 0.05) \mu_N~, \nnb \\
\mu_{\Xi_b^{\prime 0} \to \Xi_b^{0}} \es (-0.008 \pm 0.001) \mu_N~, \nnb \\
\mu_{\Xi_b^{\prime -} \to \Xi_b^{-}} \es (0.10 \pm 0.01) \mu_N~, \nnb
\eea
where upper signs correspond to the electric charge of the corresponding
negative parity baryons. 
It can easily be seen from these results that the transition magnetic moments between
the neutral $\Xi^\prime$ and $\Xi$ baryons are very close to zero. The
magnetic moments for the $\Sigma_c^+ \to \Lambda_c^+$ and $\Xi_b^{\prime +}
\to \Xi_b^+$ transitions are very close to each other which follows from
$SU(3)$ symmetry arguments.

In conclusion, the transition magnetic moments of the negative parity,
spin-1/2 heavy baryons are estimated within the QCD sum rules. The
contributions coming from the positive to positive, as well as positive to
negative parity transitions are eliminated by constructing various sum
rules. It is obtained that the magnetic moments between neutral, negative
parity heavy $\Xi_Q^{\prime 0}$ and $\Xi_Q^0$ baryons are very small.
Moreover, it is found that the magnetic moments for the $\Sigma_Q \to
\Lambda_Q$ and $\Xi_Q^{\prime \pm} \to \Xi_Q^\pm$ transitions of the
negative parity heavy baryons are quite large and can be measured in future
experiments.

\newpage


\section*{Appendix A}
In this Appendix we present the expressions of the invariant functions
$\Pi_i^B$ appearing in
the sum rules for the magnetic moment of 
$\Xi_b^{\prime 0} \to \Xi_b^0$ transition. Here in this appendix, and in
appendix B the masses of the light quarks are
neglected.\\\\

\setcounter{equation}{0}
\setcounter{section}{0}


{\bf 1) Coefficient of the $(\varepsilon\!\cdot\!p) I$ structure}

%
%
\bea
\Pi_1^B \es
- {1 \over 32 \sqrt{3} \pi^2}
(-1 + t) m_b^2 M^4 \Big\{4 (2 + t) m_b^2 \left(e_u \sp - e_s \uu\right) {\cal I}_3 \nnb \\
\ar e_b \left(\sp - \uu\right) \Big[(7 + 3 t) {\cal I}_2 - 2 (3 + t) 
m_b^2 {\cal I}_3\Big]\Big\} \nnb \\
\ar {\sqrt{3} \over 8 \pi^2}
(-1 + t) m_b^4 M^4 \left(e_s \sp - e_u \uu\right) 
   {\cal I}_3 \widetilde{j}(h_\gamma) \nnb \\
\ar {1 \over 16 \sqrt{3} \pi^2} 
 (-1 + t) (3 + t) (e_s - e_u) f_{3\gamma} m_b^3 M^4 
   \left({\cal I}_2 - m_b^2 {\cal I}_3\right) \psi^v(u_0) \nnb \\
\ar {e^{-m_b^2/M^2}\over 768 \sqrt{3} \pi^2}
(-1 + t) M^2 \Big\{12 m_0^2 \left(e_u \sp - e_s \uu\right) 
     \Big[4 + t \left(2 + m_b^2 e^{m_b^2/M^2} {\cal I}_2\right)\Big] \nnb \\
\ar e_b \left(\sp - \uu\right) 
     \Big[ 24 (7 + 3 t) m_b^2 e^{m_b^2/M^2} \left(-{\cal I}_1 + m_b^2 {\cal I}_2\right) + 
      m_0^2 \Big(7 (1 + t) + (29 + 17 t) m_b^2 e^{m_b^2/M^2} {\cal I}_2\Big)\Big]\Big\} \nnb \\
\ar {e^{-m_b^2/M^2}\over 1152 \sqrt{3} m_b \pi^2}
(-1 + t) f_{3\gamma} M^2 
\Big\{e_u \Big[-96 (1 + t) m_b \pi^2 \sp - (3 + t) \GG 
       \left(-1 + 3 m_b^2 e^{m_b^2/M^2} {\cal I}_2\right)\Big] \nnb \\
\ar e_s \Big[96 (1 + t) m_b \pi^2 \uu + 
      (3 + t) \GG \left(-1 + 3 m_b^2 e^{m_b^2/M^2} {\cal I}_2\right)\Big]\Big\} \psi^v(u_0) \nnb \\
\ar {e^{-m_b^2/M^2}\over 48 \sqrt{3} M^2}
(-1 + t^2) f_{3\gamma} m_0^2 m_b^2 \left(e_u \sp - e_s \uu\right) \psi^v(u_0) \nnb \\
\ar {e^{-m_b^2/M^2}\over 6912 \sqrt{3} M^4 \pi^2}
(-1 + t) \GG m_b^2 \left(e_u \sp - e_s \uu\right) \Big[-3 (2 + t) m_0^2 + 
   8 (1 + t) f_{3\gamma} \pi^2 \psi^v(u_0)\Big] \nnb \\
\ar {e^{-m_b^2/M^2}\over 1728 \sqrt{3} M^6}
(-1 + t^2) f_{3\gamma} \GG m_0^2 m_b^2 \left(e_u \sp - e_s \uu\right) \psi^v(u_0) \nnb \\
\ek {e^{-m_b^2/M^2}\over 3456 \sqrt{3} M^8}
(-1 + t^2) f_{3\gamma} \GG m_0^2 m_b^4 \left(e_u \sp - e_s \uu\right) \psi^v(u_0) \nnb \\
\ar {e^{-m_b^2/M^2}\over 2304 \sqrt{3} \pi^2} 
(-1 + t) \Big[4 (2 + t) \GG \left(e_u \sp - e_s \uu\right) + 
    3 (29 + 17 t) e_b m_0^2 m_b^2 e^{m_b^2/M^2} \left(\sp - \uu\right) {\cal I}_1\Big] \nnb \\
\ek {e^{-m_b^2/M^2}\over 192 \sqrt{3} \pi^2}
(-1 + t) \GG \left(e_s \sp - e_u \uu\right) 
   \widetilde{j}(h_\gamma) \nnb \\
\ek {e^{-m_b^2/M^2}\over 288 \sqrt{3}}
(-1 + t^2) f_{3\gamma} m_0^2 \left(e_u \sp - e_s \uu\right) \psi^v(u_0)~. \nnb
\eea
\\\\\\

{\bf 2) Coefficient of the $(\varepsilon \! \cdot\! p)\!\not\!p$ structure}

\bea
\Pi_2^B \es
{1\over 8 \sqrt{3} \pi^2}
(-2 + t + t^2) m_b^3 M^2 \Big[(e_b + e_u) \sp - (e_b + e_s) \uu\Big] 
  \left({\cal I}_2 - m_b^2 {\cal I}_3\right) \nnb \\
\ek {e^{-m_b^2/M^2}\over 1152 \sqrt{3} m_b M^2 \pi^2}
(-2 + t + t^2) \GG m_0^2 \left(e_u \sp - e_s \uu\right) \nnb \\
\ar {e^{-m_b^2/M^2}\over 2304 \sqrt{3} M^4 \pi^2}
(-2 + t + t^2) \GG m_0^2 m_b \left(e_u \sp - e_s \uu\right) \nnb \\
\ek {e^{-m_b^2/M^2}\over 576 \sqrt{3} m_b \pi^2}
(-1 + t) (2 + t) \GG \left(e_u \sp - e_s \uu\right) \nnb \\
\ar {\sqrt{3}\over 64 \pi^2}
(-1 + t^2) m_0^2 m_b \left(e_u \sp - e_s \uu\right) {\cal I}_1 \nnb \\
\ar {1\over 192 \sqrt{3} \pi^2}
 (-1 + t) m_b \Big\{\Big[(2 + t) e_u \GG - 
      3 (3 + 2 t) e_b m_0^2 m_b^2 -3 (7 + 5 t) e_u m_0^2 m_b^2 \Big] \sp \nnb \\
\ar \Big[-(2 + t) e_s \GG + 3 (3 + 2 t) e_b m_0^2 m_b^2 + 3 (7 + 5 t) e_s m_0^2 
       m_b^2\Big] \uu\Big\} {\cal I}_2~. \nnb
\eea
\\\\


{\bf 3) Coefficient of the $\not\!p\!\!\not\!\varepsilon$ structure}

\bea
\Pi_3^B \es
{1 \over 256 \sqrt{3} \pi^4}
(-1 + t) m_b^3 M^6 \Big[ -3 (3 + t) (e_s - e_u) 
    \left({\cal I}_2 - 2 m_b^2 {\cal I}_3 + m_b^4 {\cal I}_4\right) \nnb \\
\ar 8 (-1 + t) m_b \pi^2 
    \left(e_s \sp - e_u \uu\right) \chi \left({\cal I}_3 - m_b^2 {\cal
I}_4\right) \varphi_\gamma^\prime (u_0)\Big] \nnb \\
\ek {1 \over 768 \sqrt{3} \pi^4}
(-1 + t) (3 + t) m_b^3 M^4 \Big[ \GG (e_u - e_s) + 
    24 (e_b - e_u) m_b \pi^2 \sp \nnb \\
\ar  24 (-e_b + e_s) m_b \pi^2 \uu \Big]
   {\cal I}_3 \nnb \\
\ar {1 \over 1024 \sqrt{3} \pi^4}
(e_s - e_u) m_b M^4 \Big\{-(-1 + t) (3 + t) \GG {\cal I}_2 + 
   16 f_{3\gamma} m_b^2 \pi^2 \left(-{\cal I}_2
+ m_b^2 {\cal I}_3\right) \nnb \\ 
\cp    \Big[2 (-1 + t) (3 + t) \psi^v(u_0) - (-1 + t^2) 
      \psi^{a\prime}(u_0)\Big] \Big\} \nnb \\
\ar {1 \over 128 \sqrt{3} \pi^2}
(-1 + t) m_b^2 M^4 \left(e_s \sp - e_u \uu\right) {\cal I}_2 
   \Big\{(5 + t) i_1({\cal S},1) + (1 + 5 t) i_1(\widetilde{\cal S},1) \nnb \\
\ar 2 i_1({\cal T}_1,1) + i_1({\cal T}_2,1) + 
    2 i_1({\cal T}_3,1) - 5 i_1({\cal T}_4,1) - 6 i_1({\cal S},v)
 - 2 i_1(\widetilde{\cal S},v) \nnb \\
\ek t \Big[2 i_1({\cal T}_1,1) - 5 i_1({\cal T}_2,1) + 2 i_1({\cal T}_3,1) + i_1({\cal T}_4,1) + 2 i_1({\cal S},v) + 
      6 i_1(\widetilde{\cal S},v) \nnb \\
\ar 4 i_1({\cal T}_2,v)- 4 i_1({\cal T}_3,v)\Big] - 
    4 i_1({\cal T}_3,v) + 4 i_1({\cal T}_4,v)\Big\} \nnb \\
\ek {1 \over 128 \sqrt{3} \pi^2} 
(-1 + t) m_b^4 M^4 \left(e_s \sp - e_u \uu\right) {\cal I}_3 
   \Big\{4 (2 + t) i_1({\cal S},1) + (4 + 8 t) i_1(\widetilde{\cal S},1) \nnb \\
\ek 4 \Big[(-1 + t) i_1({\cal T}_1,1) - i_1({\cal T}_2,1) + 2 i_1({\cal T}_4,1) + 3 i_1({\cal S},v) + 
      i_1(\widetilde{\cal S},v) + i_1({\cal T}_2,v) \nnb \\
\ar t \Big(-2 i_1({\cal T}_2,1) + i_1({\cal T}_4,1) + 
        i_1({\cal S},v) + 3 i_1(\widetilde{\cal S},v)
+ i_1({\cal T}_2,v)\Big)\Big] \nnb \\
\ar 4 (1 + t) i_1({\cal T}_4,v) + 8 (2 + t) \widetilde{j}(h_\gamma) + 
    (-1 + t) \mathbb{A}^\prime (u_0)\Big\} \nnb \\
\ar {e^{-m_b^2/M^2}\over 768 \sqrt{3} \pi^2}
(-1 + t) M^2 \Big\{m_0^2 \left(e_u \sp - e_s \uu\right) \Big[-6 (3 + t) + 
      (7 + t) m_b^2 e^{m_b^2/M^2} {\cal I}_2\Big] \nnb \\
\ar e_b \left(\sp - \uu\right)
\Big[(11 + 5 t) m_0^2 - 24 (3 + t) m_b^2 e^{m_b^2/M^2} \left(-{\cal I}_1 +
m_b^2 {\cal I}_2\right)\Big]\Big\} \nnb \\
\ek {e^{-m_b^2/M^2}\over 2304 \sqrt{3} m_b \pi^2}
(-1 + t) (3 + t) f_{3\gamma} M^2 
   \Big[-(e_s - e_u) \GG - 96 m_b \pi^2 \left(e_u \sp 
- e_s \uu \right) \nnb \\
\ar 3 (e_s - e_u) \GG m_b^2 e^{m_b^2/M^2} {\cal I}_2\Big] \psi^v(u_0) \nnb \\
\ek {1 \over 2304 \sqrt{3} \pi^2}
(-1 + t)^2 \GG m_b^2 M^2 \left(e_s \sp - e_u \uu\right) \chi {\cal I}_2 
   \varphi_\gamma^\prime (u_0) \nnb \\
\ar {e^{-m_b^2/M^2}\over 4608 \sqrt{3} m_b \pi^2}
f_{3\gamma} M^2 \Big[96 t (-1 + t) m_b \pi^2 \left(e_u \sp -
e_s \uu\right) \nnb \\
\ar (-1 + t^2) (e_s - e_u) \GG \left(-1 + 3 m_b^2 e^{m_b^2/M^2} {\cal I}_2\right)\Big] 
  \psi^{a\prime}(u_0) \nnb \\
\ek {e^{-m_b^2/M^2}\over 192 \sqrt{3} M^2}
(-1 + t) f_{3\gamma} m_0^2 m_b^2 \left(e_u \sp - e_s \uu\right) 
   \Big[2 (3 + t) \psi^v(u_0) + t \psi^{a\prime}(u_0)\Big] \nnb \\
\ar {e^{-m_b^2/M^2}\over 27648 \sqrt{3} M^4 \pi^2}
(-1 + t) \GG m_b^2 \left(e_u \sp - e_s \uu\right) \Big\{3 (3 + t) m_0^2 \nnb \\
\ek 8 f_{3\gamma} \pi^2 \Big[2 (3 + t) \psi^v(u_0) + 
     t \psi^{a\prime}(u_0)\Big]\Big\} \nnb \\
\ek {e^{-m_b^2/M^2}\over 6912 \sqrt{3} M^6}
(-1 + t) f_{3\gamma} \GG m_0^2 m_b^2 \left(e_u \sp - e_s \uu\right) 
   \Big[2 (3 + t) \psi^v(u_0) + t \psi^{a\prime}(u_0)\Big] \nnb \\
\ar {e^{-m_b^2/M^2}\over 13824 \sqrt{3} M^8}
(-1 + t) f_{3\gamma} \GG m_0^2 m_b^4 \left(e_u \sp - e_s \uu\right) 
  \Big[2 (3 + t) \psi^v(u_0) + t \psi^{a\prime}(u_0)\Big] \nnb \\
\ar {e^{-m_b^2/M^2}\over 9216 \sqrt{3} \pi^2}
(-1 + t) \GG \left(e_s \sp - e_u \uu\right) \Big\{3 (1 + t) i_1({\cal S},1) + 
   3 (1 + t) i_1(\widetilde{\cal S},1) \nnb \\
\ar 2 i_1({\cal T}_1,1) + 3 i_1({\cal T}_2,1) - 2 i_1({\cal T}_3,1) - 
   3 i_1({\cal T}_4,1) - 6 i_1({\cal S},v) - 2 i_1(\widetilde{\cal S},v) - 4 i_1({\cal T}_2,v) \nnb \\
\ar 4 i_1({\cal T}_3,v) + 16 \widetilde{j}(h_\gamma) - \mathbb{A}^\prime (u_0) + 
   t \Big[-2 i_1({\cal T}_1,1) + 3 i_1({\cal T}_2,1) + 2 i_1({\cal T}_3,1) - 3 i_1({\cal T}_4,1) \nnb \\
\ek 2 i_1({\cal S},v) - 6 i_1(\widetilde{\cal S},v) - 4 i_1({\cal T}_3,v) + 4 i_1({\cal T}_4,v) + 
     8 \widetilde{j}(h_\gamma) + \mathbb{A}^\prime (u_0)\Big]\Big\} \nnb \\
\ek {e^{-m_b^2/M^2}\over 2304 \sqrt{3} \pi^2}
(-1 + t) \Big\{(3 + t) \GG \left(e_u \sp - e_s \uu\right) + 
    3 (11 + 5 t) e_b m_0^2 m_b^2 e^{m_b^2/M^2} \left(\sp - \uu\right) {\cal I}_1 \nnb \\
\ar 2 f_{3\gamma} m_0^2 \pi^2 \left(e_u \sp - e_s \uu\right) \Big[2 (11 + 5 t) \psi^v(u_0) + 
      (2 + 5 t) \psi^{a\prime}(u_0)\Big]\Big\}~. \nnb
\eea
\\\\\\


{\bf 4) Coefficient of the $\not\!\varepsilon$ structure}

\bea
\Pi_4^B \es
{\sqrt{3}\over 32 \pi^4}
(1 + t + t^2) (e_s - e_u) m_b^4 M^8 
  \left(-{\cal I}_3 + m_b^2 {\cal I}_4\right) \nnb \\
\ar {1\over 128 \sqrt{3} \pi^2}
(e_s - e_u) f_{3\gamma} m_b^2 M^6 \left[-3 (1 + t)^2 {\cal I}_2 + 
    4 (1 + t + t^2) m_b^2 {\cal I}_3\right] i_2({\cal A},v) \nnb \\
\ar {1\over 128 \sqrt{3} \pi^2}
(e_s - e_u) f_{3\gamma} m_b^2 M^6 \left[-3 (1 + t)^2 {\cal I}_2 + 
    2 (1 + 4 t + t^2) m_b^2 {\cal I}_3\right] i_2({\cal V},v) \nnb \\
\ek {1\over 32 \sqrt{3} \pi^2}
(1 + t + t^2) (e_s - e_u) f_{3\gamma} m_b^4 M^6 {\cal I}_3
\left[4 \psi^v(u_0)-\psi^{a\prime}(u_0)\right] \nnb \\
\ar {\sqrt{3}\over 64 \pi^2}
(-1 + t^2) m_b^3 M^6 \left (e_s \sp - e_u \uu\right) \chi 
   \left({\cal I}_2 - m_b^2 {\cal I}_3\right) \varphi_\gamma^\prime (u_0) \nnb \\
\ar {e^{-m_b^2/M^2}\over 1536 \sqrt{3} m_b \pi^2}
(-1 + t^2) \GG M^4 \left(e_s \sp - e_u \uu\right)
\chi \left(-1 + 3 m_b^2 e^{m_b^2/M^2} {\cal I}_2\right)
\varphi_\gamma^\prime (u_0) \nnb \\
\ek {1 \over 16 \sqrt{3} \pi^2}
(-2 + t + t^2) m_b^5 M^4 \left[(e_b + e_u) \sp - (e_b + e_s)
\uu\right] {\cal I}_3 \nnb \\
\ek {\sqrt{3} \over 128 \pi^2}
(-1 + t^2) m_b M^4 \left(e_s \sp - e_u \uu\right) {\cal I}_1 
    \left[i_1({\cal T}_1,1) + i_1({\cal T}_3,1)\right] \nnb \\
\ek {1 \over 128 \sqrt{3} \pi^2}
(-1 + t) m_b M^4 \left(e_s \sp - e_u \uu\right) {\cal I}_1 
   \Big\{(5 + t) i_1({\cal S},1) \nnb \\
\ek (1 + 5 t) \left[i_1(\widetilde{\cal S},1) + i_1({\cal T}_2,1)\right] - 
    (5 + t) i_1({\cal T}_4,1)\Big\} \nnb \\
\ek {1 \over 768 \sqrt{3} \pi^4}
m_b^2 M^4 {\cal I}_2 \Big\{-(1 + t + t^2) e_s \GG + 
    (1 + t + t^2) e_u \GG \nnb \\
\ek 48 (-2 + t + t^2) e_b m_b \pi^2 
     \left(\sp - \uu\right) + 3 (-1 + t) m_b \pi^2 \left(e_s \sp - e_u \uu\right) \nnb \\
\cp \Big[2 (-1 + 7 t) \left(i_1(\widetilde{\cal S},1) + i_1({\cal T}_2,1)\right) + 
      2 (-7 + t) \left(i_1({\cal S},1) - i_1({\cal T}_4,1)\right) \nnb\\
\ar 8 (2 + t) i_1({\cal S},v) - 
      4 (-1 + t) i_1(\widetilde{\cal S},v) + 8 t i_1({\cal T}_4,v) + 
      4 (3 + t) \left(i_1({\cal T}_2,v) - 2 \widetilde{j}(h_\gamma)\right) \nnb \\
\ar 3 (1 + t) \left(-4 (i_1({\cal T}_1,1) + i_1({\cal T}_3,v)) + \mathbb{A}^\prime
      (u_0)\right)\Big]\Big\} \nnb \\
\ek {e^{-m_b^2/M^2}\over 9216 \sqrt{3} \pi^2}
(-1 + t)^2 (e_s - e_u) f_{3\gamma} \GG M^2 \left[i_2({\cal A},v) -
i_2({\cal V},v)\right] \nnb \\
\ek {e^{-m_b^2/M^2}\over 9216 \sqrt{3} m_b \pi^2}
(-1 + t) \GG M^2 \left(e_s \sp - e_u \uu\right) \Big\{4 (-1 + t) i_1({\cal S},1) \nnb \\
\ar 2 \Big[2 (-1 + t) i_1(\widetilde{\cal S},1) - 3 (1 + t) i_1({\cal T}_1,1) - 2 i_1({\cal T}_2,1) + 
      3 i_1({\cal T}_3,1) \nnb \\
\ar  2 \Big(i_1({\cal T}_4,1) + 4 i_1({\cal S},v) + i_1(\widetilde{\cal S},v) + 
        3 i_1({\cal T}_2,v) - 3 i_1({\cal T}_3,v) - 6 \widetilde{j}(h_\gamma) \Big) \nnb \\
\ar t \Big(2 i_1({\cal T}_2,1) + 3 i_1({\cal T}_3,1) - 2 i_1({\cal T}_4,1)
+ 4 i_1({\cal S},v) - 2 i_1(\widetilde{\cal S},v) + 2 i_1({\cal T}_2,v) \nnb \\
\ek 6 i_1({\cal T}_3,v) + 4  i_1({\cal T}_4,v) - 
          4 \widetilde{j}(h_\gamma)\Big)\Big] + 3 (1 + t) \mathbb{A}^\prime (u_0)\Big\} \nnb \\
\ek {1 \over 128 \sqrt{3} \pi^2}
(-1 + t) (7 + 5 t) m_0^2 m_b^3 M^2 \left(e_u \sp - e_s \uu\right) {\cal I}_2 \nnb \\
\ek {e^{-m_b^2/M^2}\over 384 \sqrt{3} m_b \pi^2}
(-2 + t + t^2) M^2 \left(e_u \sp - e_s \uu\right) \Big[\GG \left(1 -
e^{m_b^2/M^2} m_b^2 {\cal I}_2\right) \nnb \\
\ek 6 m_0^2 m_b^2 + 32 f_{3\gamma} m_b^2 \pi^2 \psi^v(u_0)\Big] \nnb \\
\ar {e^{-m_b^2/M^2}\over 2304 \sqrt{3} \pi^2}
(1 + t + t^2) (e_s - e_u) f_{3\gamma} \GG M^2 
\left[4 \psi^v(u_0) - \psi^{a\prime}(u_0)\right] \nnb \\
\ar {e^{-m_b^2/M^2}\over 48 \sqrt{3}}
(1 + t - 2 t^2) e_s f_{3\gamma} m_b M^2 \uu \psi^{a\prime}(u_0) \nnb \\
\ar {e^{-m_b^2/M^2}\over 384 \sqrt{3} \pi^2} 
(-1 + t) m_b M^2 \Big[- e^{m_b^2/M^2}3 (3 + 2 t) e_b m_0^2 m_b^2 
\left(\sp - \uu\right) {\cal I}_2 \nnb \\
\ar 8 (1 + 2 t) e_u f_{3\gamma} \pi^2 \sp \psi^{a\prime}(u_0)\Big] \nnb \\
\ar {e^{-m_b^2/M^2}\over 1152 \sqrt{3} M^2 \pi^2}
(-1 + t) m_b \left(e_u \sp - e_s \uu\right) \Big\{(2 + t) \GG m_0^2 \nnb \\
\ar f_{3\gamma} \left(\GG - 6 m_0^2 m_b^2\right) \pi^2 \Big[-4 (2 + t) \psi^v(u_0) + 
     (1 + 2 t) \psi^{a\prime}(u_0)\Big]\Big\} \nnb \\
\ek {e^{-m_b^2/M^2}\over 13824 \sqrt{3} M^4 \pi^2}
(-1 + t) \GG m_b \left(e_u \sp - e_s \uu\right) \Big\{3 (2 + t) m_0^2 m_b^2 \nnb \\
\ar    2 f_{3\gamma} (3 m_0^2 - 2 m_b^2) \pi^2 \Big[4 (2 + t) \psi^v(u_0) - 
      (1 + 2 t) \psi^{a\prime}(u_0)\Big]\Big\} \nnb \\
\ar {e^{-m_b^2/M^2}\over 2304 \sqrt{3} M^6}
(-1 + t) f_{3\gamma} \GG m_0^2 m_b^3 \left(e_u \sp - e_s \uu\right) 
  \Big[4 (2 + t) \psi^v(u_0) - (1 + 2 t) \psi^{a\prime}(u_0)\Big] \nnb \\
\ek {e^{-m_b^2/M^2}\over 13824 \sqrt{3} M^8}
(-1 + t) f_{3\gamma} \GG m_0^2 m_b^5 \left(e_u \sp - e_s \uu\right) 
   \Big[4 (2 + t) \psi^v(u_0) - (1 + 2 t) \psi^{a\prime}(u_0)\Big] \nnb \\
\ar {e^{-m_b^2/M^2}\over 18432 \sqrt{3} \pi^2}
(-1 + t) \GG m_b \left(e_s \sp - e_u \uu\right) \Big\{4 (-1 + t) i_1({\cal S},1) \nnb \\
\ar 2 \Big[2 (-1 + t) i_1(\widetilde{\cal S},1) - 3 (1 + t) i_1({\cal T}_1,1)
- 2 i_1({\cal T}_2,1) + 3 i_1({\cal T}_3,1) + 2 i_1({\cal T}_4,1) +
8 i_1({\cal S},v) \nnb \\
\ar 2 i_1(\widetilde{\cal S},v) + 6 i_1({\cal T}_2,v) - 6 i_1({\cal T}_3,v)
- 12 \widetilde{j}(h_\gamma) + 
t \Big(2 i_1({\cal T}_2,1) + 3 i_1({\cal T}_3,1) - 2 i_1({\cal T}_4,1)\nnb \\
\ar 4 i_1({\cal S},v) - 2 i_1(\widetilde{\cal S},v) + 2 i_1({\cal T}_2,v)
- 6 i_1({\cal T}_3,v) + 4 i_1({\cal T}_4,v) - 4 \widetilde{j}(h_\gamma)\Big)\Big]
+ 3 (1 + t) \mathbb{A}^\prime (u_0)\Big\} \nnb \\
\ar {e^{-m_b^2/M^2}\over 2304 \sqrt{3} m_b \pi^2} 
\left(e_u \sp - e_s \uu\right) \Big\{-(-2 + t + t^2) \GG (m_0^2 - 2 m_b^2) \nnb \\
\ar 18 (-1 + t^2) f_{3\gamma} m_0^2 m_b^2 \pi^2 \Big[-4 \psi^v(u_0) + 
      \psi^{a\prime}(u_0)\Big]\Big\}~. \nnb
\eea


The functions $i_n~(n=1,2)$, and $\widetilde{j}_1(f(u))$
are defined as:
\bea
\label{nolabel}
i_1(\phi,f(v)) \es \int {\cal D}\alpha_i \int_0^1 dv 
\phi(\alpha_{\bar{q}},\alpha_q,\alpha_g) f(v) \delta^\prime(k-u_0)~, \nnb \\
i_2(\phi,f(v)) \es \int {\cal D}\alpha_i \int_0^1 dv 
\phi(\alpha_{\bar{q}},\alpha_q,\alpha_g) f(v) \delta^{\prime\prime}(k-u_0)~, \nnb \\
\widetilde{j}(f(u)) \es \int_{u_0}^1 du f(u)~, \nnb \\
{\cal I}_n \es \int_{m_b^2}^{\infty} ds\, {e^{-s/M^2} \over s^n}~,\nnb
\eea
where 
\bea
k = \alpha_q + \alpha_g \bar{v}~,~~~~~u_0={M_1^2 \over M_1^2
+M_2^2}~,~~~~~M^2={M_1^2 M_2^2 \over M_1^2 +M_2^2}~.\nnb
\eea




\newpage

\section*{Appendix B}

Expressions of the invariant amplitudes $\Pi_1^M$ and
$\Pi_2^M$ entering into the mass sum rule for the negative parity
heavy $\Xi_b^{\prime 0}$ baryon. \\\\

\setcounter{equation}{0}
\setcounter{section}{0}


{\bf 1) Coefficient of the $\not\!p$ structure}

%
%
\bea
&&\Pi_1^M =
{3\over 256 \pi^4} \Big\{ - m_b^4  M^6  [5 + t (2 + 5 t)] \left[
m_b^4 {\cal I}_5 -
2 m_b^2 {\cal I}_4 +
{\cal I}_3\right] \Big\} \nnb \\
\ar {1\over 192 \pi^4} m_b^4 M^2 \left[\GG  (1 + t + t^2 ) - 18 m_b \pi^2 (-1+t^2)
\left(\sp+\uu\right)\right] {\cal I}_3 \nnb \\
\ar {1\over 3072 \pi^4} m_b^2 M^2 \left[-\GG  (13 + 10 t + 13 t^2 ) + 288 m_b \pi^2
(-1+t^2) \left(\sp+\uu\right)\right] {\cal I}_2 \nnb \\
\ar {e^{-m_b^2/M^2} \over 73728 m_b M^2 \pi^4} \Big\{ - \GG^2 m_b (1 + t)^2 +
768 m_b m_0^2 \sp \uu \pi^4 (-1 + t)^2 \nnb \\
\ek 56 \GG m_0^2 \pi^2 (-1 + t^2) \left(\sp + \uu\right) \Big\} \nnb \\
\ar {1 \over 768 M^2 \pi^2} \GG m_b (-1 + t^2) (\sp + \uu) {\cal  E}_1 \nnb \\
\ar {e^{-m_b^2/M^2} \over 18432 M^4  \pi^2} m_b m_0^2
\left[\GG \left(\uu+\sp\right) (-1+t^2) +
384 m_b \sp \uu \pi^2 (-1+t)^2 \right] \nnb \\
\ar {e^{-m_b^2/M^2} \over 1728 M^6} m_b^2 \GG (-1 + t)^2 \sp \uu \nnb \\
\ar {e^{-m_b^2/M^2} \over 1728 M^8} m_b^2 m_0^2 \GG (-1 + t)^2 \sp \uu \nnb \\
\ek {e^{-m_b^2/M^2} \over 3456 M^{10}} m_b^4 m_0^2 \GG (-1 + t)^2 \sp \uu \nnb \\
\ek {e^{-m_b^2/M^2} \over 768 m_b \pi^2} \left[ \GG \left(\uu+\sp\right)
(-1+t^2) + 32 m_b \sp \uu \pi^2 (-1+t)^2 \right] \nnb \\
\ar {1\over 256 \pi^2} m_b \left(\uu+\sp\right) (-1+t^2) \left[
\left(\GG - 13 m_b^2 m_0^2\right) {\cal I}_2 + 6 m_0^2 {\cal I}_1
\right]~,\nnb
\eea
\\\\
{\bf 2) Coefficient of the $I$ structure}
%
%
\bea
&&\Pi_2^M =
-{3\over 256 \pi^4} \Big\{ - m_b^3  M^6 (-1+t)^2 \left[          
m_b^4 {\cal I}_4 -                          
2 m_b^2 {\cal I}_3 +                               
{\cal I}_2\right] \Big\} \nnb \\       
\ar {1\over 3072 \pi^4} m_b M^4 \Big\{4 m_b^2 \left[\GG (-1+t)^2 + 72 m_b
\left(\uu+\sp\right) \pi^2 (-1+t^2) \right] {\cal I}_3
- 3 \GG (-1+t)^2 {\cal I}_2 \Big\} \nnb \\
\ek {7 e^{-m_b^2/M^2} \over 256 \pi^2} m_0^2 M^2
\left(\uu+\sp\right) (-1 + t^2) \nnb \\
\ar {1\over 1024 \pi^4} m_b M^2 \Big\{
m_b \left[ 3 m_b \GG (-1+t)^2 + 4 m_0^2 \left(\uu+\sp\right) \pi^2 (-1+t^2)
\right] {\cal I}_2 -
2 \GG (-1+t)^2 {\cal I}_1 \Big\} \nnb \\
\ek {e^{-m_b^2/M^2} \over 73728 M^2 \pi^4} m_b 
\left[ \GG^2 (-1 + t)^2 +
1536 m_0^2 \sp \uu \pi^4 (3 + 2 t + 3 t^2) \right] \nnb \\
\ar {e^{-m_b^2/M^2} \over 18432 M^4  \pi^2} m_b 
\left[ -11 m_b m_0^2 \GG \left(\uu+\sp\right) (-1+t^2)\right. \nnb \\
\ek \left. 32 \left(\GG - 12 m_0^2 m_b^2 \right) \sp \uu (5 + 2 t + 5 t^2) 
\right] \nnb \\
\ar {e^{-m_b^2/M^2} \over 1728 M^6} m_b \left( m_b^2 - 3 m_0^2 \right)
\GG \sp \uu (5 + 2 t + 5 t^2) \nnb \\
\ar {e^{-m_b^2/M^2} \over 576 M^8} m_b^3 m_0^2 \GG \sp \uu (5 + 2 t + 5 t^2)
\nnb \\
\ek {e^{-m_b^2/M^2} \over 3456 M^{10}} m_b^5 m_0^2 \GG \sp \uu (5 + 2 t + 5
t^2) \nnb \\
\ar {e^{-m_b^2/M^2} \over 36864 m_b \pi^4} \left[ \GG^2 (-1+t)^2
- 1536 m_b^2 \sp \uu \pi^4 (5 + 2 t + 5 t^2) \right. \nnb \\
\ar \left.96 m_b \GG \left(\uu+\sp\right) \pi^2 (-1+t^2) \right]~. \nnb
\eea
\\\\


Expressions of the invariant amplitudes $\Pi_1^M$ and
$\Pi_2^M$ entering into the mass sum rule for the negative parity
heavy $\Xi_b^0$ baryon.\\\\

{\bf 3) Coefficient of the $\not\!p$ structure}

%
%
\bea
&&\Pi_1^M =
- {1\over 256 \pi^4}
(-3 m_b^4 M^6 (5 + 2 t + 5 t^2) \left({\cal I}_3 - 2 m_b^2 {\cal I}_4 + m_b^4 {\cal I}_5\right) \nnb \\
\ar {1\over 3072 \pi^4}
m_b^2 M^2 \Big[3 \GG (1 + t)^2 {\cal I}_2 - 16 \GG m_b^2 (1 + t + t^2) {\cal I}_3 \nnb \\
\ek 32 m_b \pi^2 (-1 + t) (1 + 5 t) \left(\sp + \uu\right) 
\left(-{\cal I}_2 + m_b^2 {\cal I}_3\right)\Big] \nnb \\
\ar {e^{-m_b^2/M^2} \over 221184 m_b M^2 \pi^4} 
\GG^2 m_b (13 + 10 t + 13 t^2) + 768 m_0^2 m_b \pi^4 (-1 + t) (25 + 23 t) \sp \uu \nnb \\
\ek 8 \GG \pi^2 (-1 + t) (\sp + \uu) \left[m_0^2 (1 + 5 t) + 
    12 m_b^2 e^{m_b^2/M^2} (5 + t) {\cal I}_1\right] \nnb \\
\ar {e^{-m_b^2/M^2} \over 55296 M^4 \pi^2}
m_0^2 m_b (-1 + t) \left[384 m_b \pi^2 (13 + 11 t) \sp \uu + 
   \GG (31 + 11 t) \left(\sp + \uu\right)\right] \nnb \\
\ar {e^{-m_b^2/M^2} \over 5184 M^6}
\GG m_b^2 (-1 + t) (13 + 11 t) \sp \uu \nnb \\
\ar {e^{-m_b^2/M^2} \over 5184 M^8}
\GG m_0^2 m_b^2 (-1 + t) (13 + 11 t) \sp \uu \nnb \\
\ek {e^{-m_b^2/M^2} \over 10368 M^{10}}
\GG m_0^2 m_b^4 (-1 + t) (13 + 11 t) \sp \uu \nnb \\
\ar {e^{-m_b^2/M^2} \over 6912 m_b \pi^2}
(-1 + t) \Big\{\GG (1 + 5 t) \left(\sp + \uu\right) 
\left(-1 + 3 m_b^2 e^{m_b^2/M^2} {\cal I}_2\right) \nnb \\
\ek 3 m_b \Big[32 \pi^2 (13 + 11 t) \sp \uu + 3 m_0^2 m_b e^{m_b^2/M^2} \left(\sp + \uu\right) 
      [-6 (1 + t) {\cal I}_1 + m_b^2 (7 + 11 t) 
{\cal I}_2]\Big]\Big\}~.\nnb
\eea

{\bf 4) Coefficient of the $I$ structure}
%
%
\bea
&&\Pi_2^M =
- {1\over 256 \pi^4}
m_b^3 M^6 (-1 + t) (13 + 11 t) \left({\cal I}_2 - 2 m_b^2 {\cal I}_3 +
m_b^4 {\cal I}_4\right) \nnb \\
\ek {1\over 9216 \pi^4}
m_b M^4 (-1 + t) \Big[-96 m_b^3 \pi^2 (1 + 5 t) (\sp + \uu) {\cal I}_3 + 
    \GG (13 + 11 t) \left(3 {\cal I}_2 - 4 m_b^2 {\cal I}_3\right)\Big] \nnb \\
\ek {e^{-m_b^2/M^2} \over 3072 \pi^4}
M^2 (-1 + t) \Big\{4 m_0^2 \pi^2 \left(\sp + \uu\right) \left[1 + 5 t +
m_b^2 e^{m_b^2/M^2} (5 + t) {\cal I}_2\right] \nnb \\
\ar  \GG m_b e^{m_b^2/M^2} \left[-2 (-1 + t) {\cal I}_1 +
3 m_b^2 (3 + 5 t) {\cal I}_2\right]\Big\} \nnb \\
\ar {e^{-m_b^2/M^2} \over 221184 M^2 \pi^4}
m_b (-1 + t) \Big[\GG^2 (11 + 13 t) - 1536 m_0^2 \pi^4 (-1 + t) \sp \uu \Big] \nnb \\
\ar {e^{-m_b^2/M^2} \over 55296 M^4 \pi^2}
m_b \Big\{1152 m_0^2 m_b^2 \pi^2 (5 + 2 t + 5 t^2) \sp \uu \nnb \\
\ek \GG \Big[96 \pi^2 (5 + 2 t + 5 t^2) \sp \uu + m_0^2 m_b (-1 + t) (29 + t) 
      \left(\sp + \uu\right)\Big]\Big\} \nnb \\
\ar {e^{-m_b^2/M^2} \over 1728 M^6}
\GG m_b (-3 m_0^2 + m_b^2) (5 + 2 t + 5 t^2) \sp \uu \nnb \\
\ar {e^{-m_b^2/M^2} \over 576 M^8}
\GG m_0^2 m_b^3 (5 + 2 t + 5 t^2) \sp \uu \nnb \\
\ek {e^{-m_b^2/M^2} \over 3456 M^{10}}
\GG m_0^2 m_b^5 (5 + 2 t + 5 t^2) \sp \uu \nnb \\
\ar {e^{-m_b^2/M^2} \over 110592 m_b \pi^4}
\Big[\GG^2 (11 + 2 t - 13 t^2) - 4608 m_b^2 \pi^4 (5 + 2 t + 5 t^2) \sp \uu \nnb \\
\ek 32 \GG m_b \pi^2 (-7 + t) (-1 + t) \left(\sp + \uu\right)\Big]~.\nnb
\eea
where
\bea
{\cal I}_n = \int_{m_b^2}^{\infty} ds\, {e^{-s/M^2} \over s^n}~.\nnb
\eea




\newpage

\section*{Appendix C: Photon distribution amplitudes}
\setcounter{equation}{0}
\setcounter{section}{0}


Explicit forms of the photon DAs
\cite{Rfrd18}.

\bea
\label{nolabel27}
\varphi_\gamma(u) \es 6 u \bar u \Big[ 1 + \varphi_2(\mu)
C_2^{\frac{3}{2}}(u - \bar u) \Big]~,
\nnb \\
\psi^v(u) \es 3 [3 (2 u - 1)^2 -1 ]+\frac{3}{64} (15
w^V_\gamma - 5 w^A_\gamma)
                        [3 - 30 (2 u - 1)^2 + 35 (2 u -1)^4]~,
\nnb \\
\psi^a(u) \es [1- (2 u -1)^2] [ 5 (2 u -1)^2 -1 ]
\frac{5}{2}
    \Bigg(1 + \frac{9}{16} w^V_\gamma - \frac{3}{16} w^A_\gamma
    \Bigg)~,
\nnb \\
{\cal A}(\alpha_i) \es 360 \alpha_q \alpha_{\bar q} \alpha_g^2
        \Bigg[ 1 + w^A_\gamma \frac{1}{2} (7 \alpha_g - 3)\Bigg]~,
\nnb \\
{\cal V}(\alpha_i) \es 540 w^V_\gamma (\alpha_q - \alpha_{\bar q})
\alpha_q \alpha_{\bar q}
                \alpha_g^2~,
\nnb \\
h_\gamma(u) \es - 10 (1 + 2 \kappa^+ ) C_2^{\frac{1}{2}}(u
- \bar u)~,
\nnb \\
\mathbb{A}(u) \es 40 u^2 \bar u^2 (3 \kappa - \kappa^+ +1 ) +
        8 (\zeta_2^+ - 3 \zeta_2) [u \bar u (2 + 13 u \bar u) + 
                2 u^3 (10 -15 u + 6 u^2) \ln(u) \nnb \\ 
\ar 2 \bar u^3 (10 - 15 \bar u + 6 \bar u^2)
        \ln(\bar u) ]~,
\nnb \\
{\cal T}_1(\alpha_i) \es -120 (3 \zeta_2 + \zeta_2^+)(\alpha_{\bar
q} - \alpha_q)
        \alpha_{\bar q} \alpha_q \alpha_g~,
\nnb \\
{\cal T}_2(\alpha_i) \es 30 \alpha_g^2 (\alpha_{\bar q} - \alpha_q)
    [(\kappa - \kappa^+) + (\zeta_1 - \zeta_1^+)(1 - 2\alpha_g) +
    \zeta_2 (3 - 4 \alpha_g)]~,
\nnb \\
{\cal T}_3(\alpha_i) \es - 120 (3 \zeta_2 - \zeta_2^+)(\alpha_{\bar
q} -\alpha_q)
        \alpha_{\bar q} \alpha_q \alpha_g~,
\nnb \\
{\cal T}_4(\alpha_i) \es 30 \alpha_g^2 (\alpha_{\bar q} - \alpha_q)
    [(\kappa + \kappa^+) + (\zeta_1 + \zeta_1^+)(1 - 2\alpha_g) +
    \zeta_2 (3 - 4 \alpha_g)]~,\nnb \\
{\cal S}(\alpha_i) \es 30\alpha_g^2\{(\kappa +
\kappa^+)(1-\alpha_g)+(\zeta_1 + \zeta_1^+)(1 - \alpha_g)(1 -
2\alpha_g)\nnb \\ 
\ar\zeta_2
[3 (\alpha_{\bar q} - \alpha_q)^2-\alpha_g(1 - \alpha_g)]\}~,\nnb \\
\widetilde {\cal S}(\alpha_i) \es-30\alpha_g^2\{(\kappa -
\kappa^+)(1-\alpha_g)+(\zeta_1 - \zeta_1^+)(1 - \alpha_g)(1 -
2\alpha_g)\nnb \\ 
\ar\zeta_2 [3 (\alpha_{\bar q} -
\alpha_q)^2-\alpha_g(1 - \alpha_g)]\}. \nnb
\eea
The parameters entering  the above DA's are borrowed from
\cite{Rfrd18} whose values are $\varphi_2(1~GeV) = 0$, 
$w^V_\gamma = 3.8 \pm 1.8$, $w^A_\gamma = -2.1 \pm 1.0$, 
$\kappa = 0.2$, $\kappa^+ = 0$, $\zeta_1 = 0.4$, $\zeta_2 = 0.3$, 
$\zeta_1^+ = 0$, and $\zeta_2^+ = 0$.


\newpage


\newpage

\section*{Figure captions}
{\bf Fig. (1)} The dependence of the magnetic moment of the negative parity
$\Sigma_0^b \to \Lambda_b^0$ transition 
on $M^2$, at several fixed values of $t$, and at
$s_0=40.0~GeV^2$, in units of nuclear magneton $\mu_N$.\\\\
{\bf Fig. (2)} The same as Fig. (1), but at $s_0=42.5~GeV^2$.\\\\
{\bf Fig. (3)} The dependence of the magnetic moment of the negative parity
$\Sigma_0^b \to \Lambda_b^0$ transition
on $\cos\theta$, at several fixed values of $M^2$, and at
$s_0=40.0~GeV^2$, in units of nuclear magneton $\mu_N$.\\\\
{\bf Fig. (4)} The same as Fig. (3), but at $s_0=42.5~GeV^2$.

\newpage

\begin{figure}
\vskip 3. cm
    \includegraphics{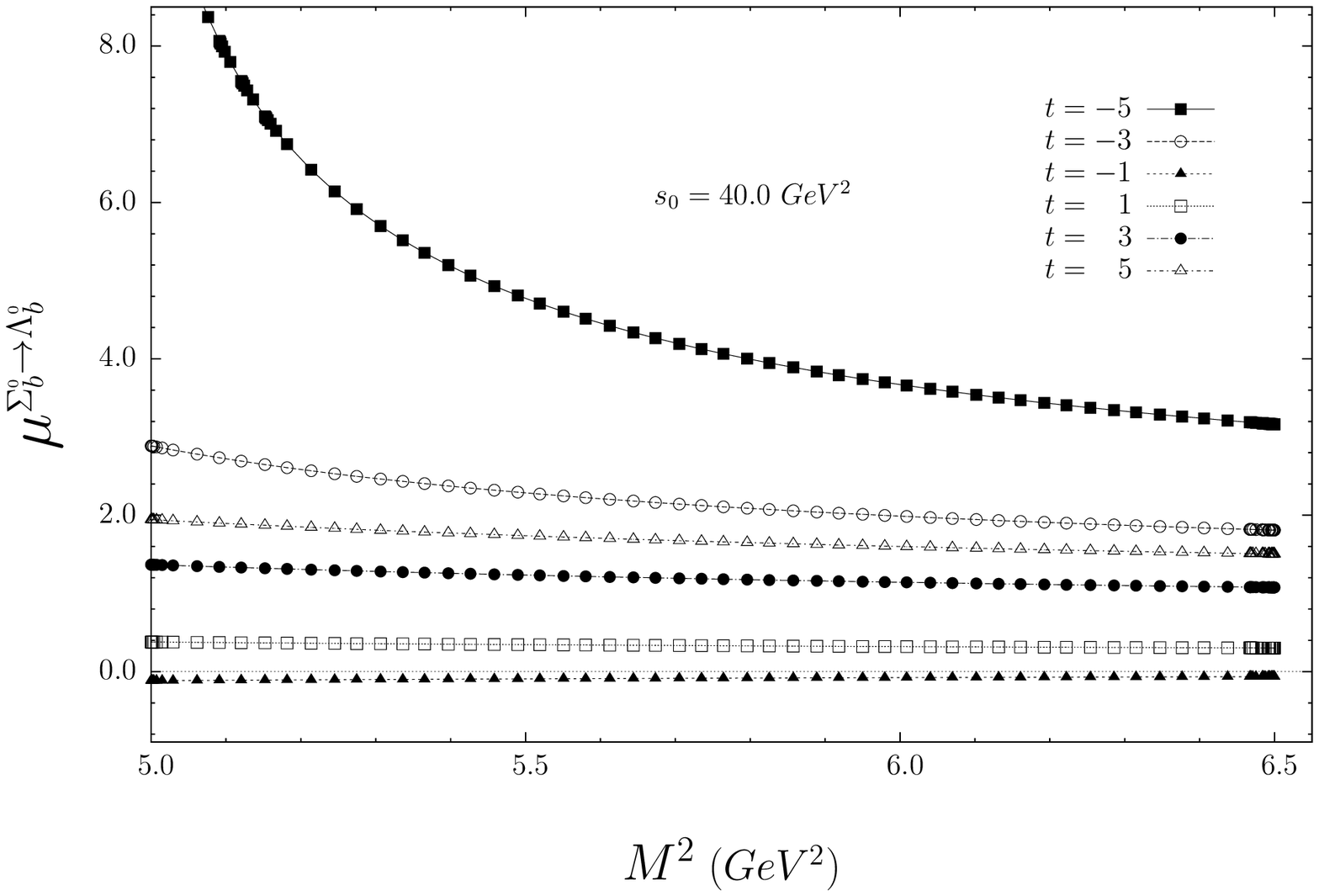}
\vskip 7.0cm
\caption{}
\end{figure}

\begin{figure}
\vskip 3. cm
    \includegraphics{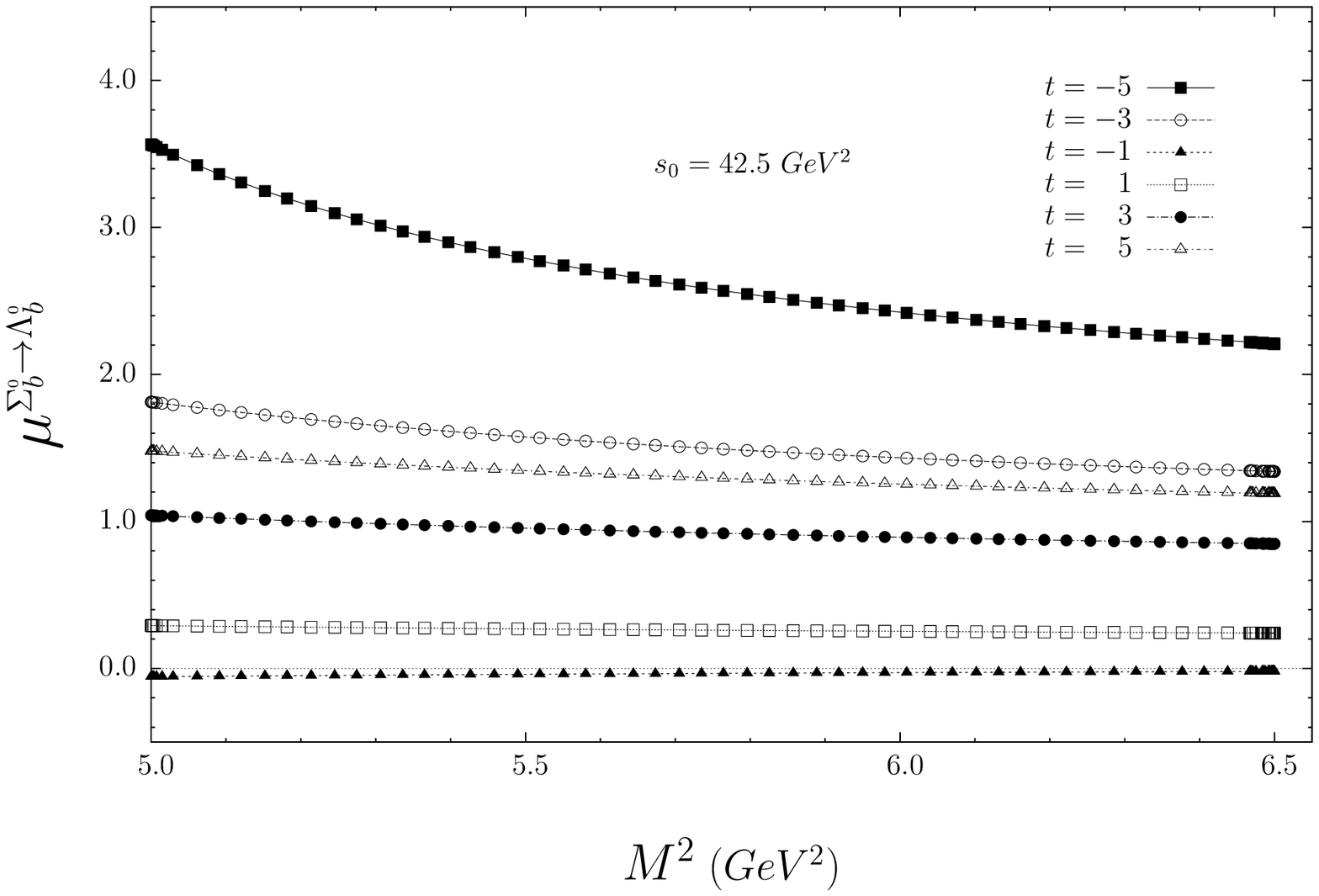}
\vskip 7.0cm
\caption{}
\end{figure}

\begin{figure}
\vskip 3. cm
    \includegraphics{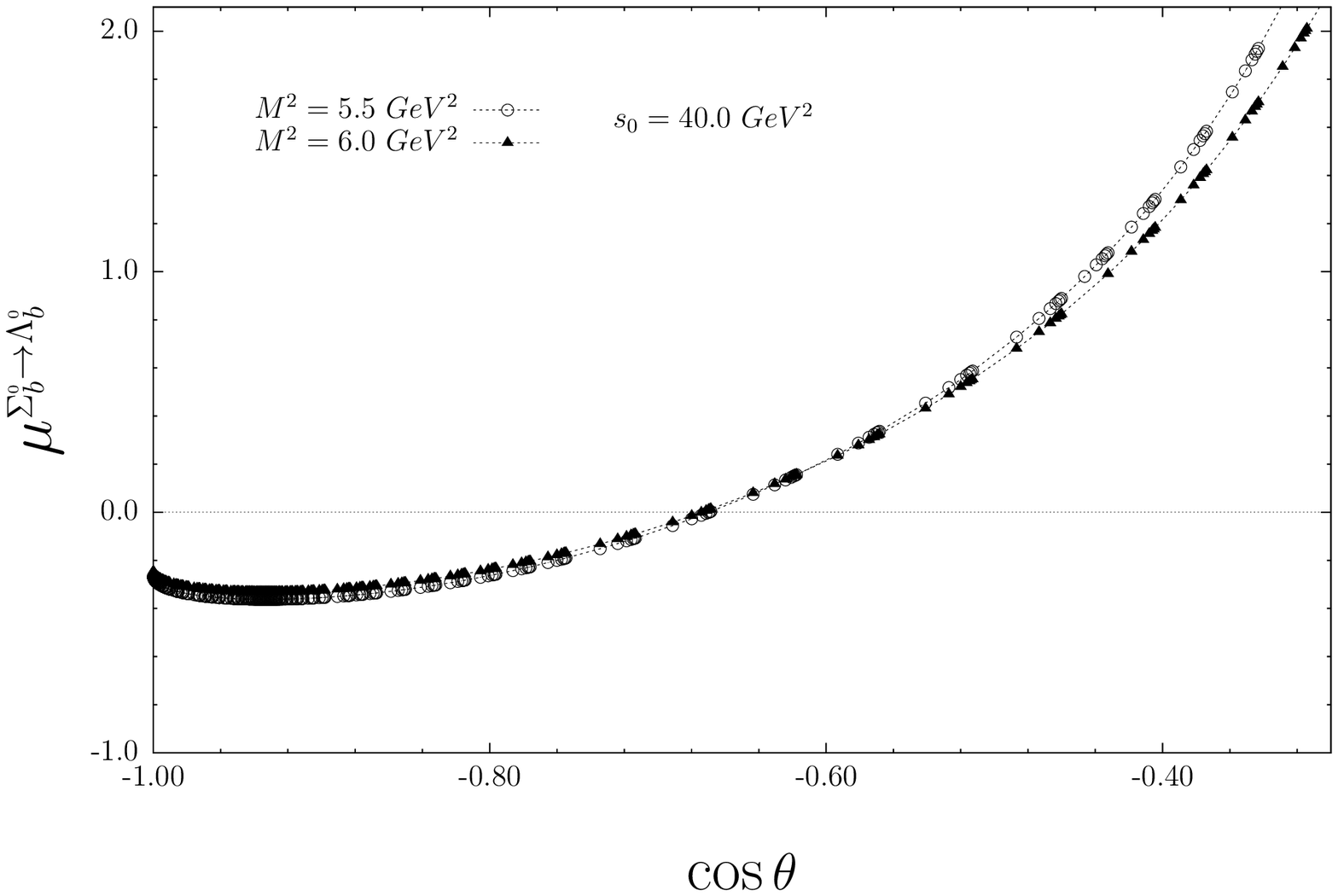}
\vskip 7.0cm
\caption{}
\end{figure}

\begin{figure}
\vskip 3. cm
    \includegraphics{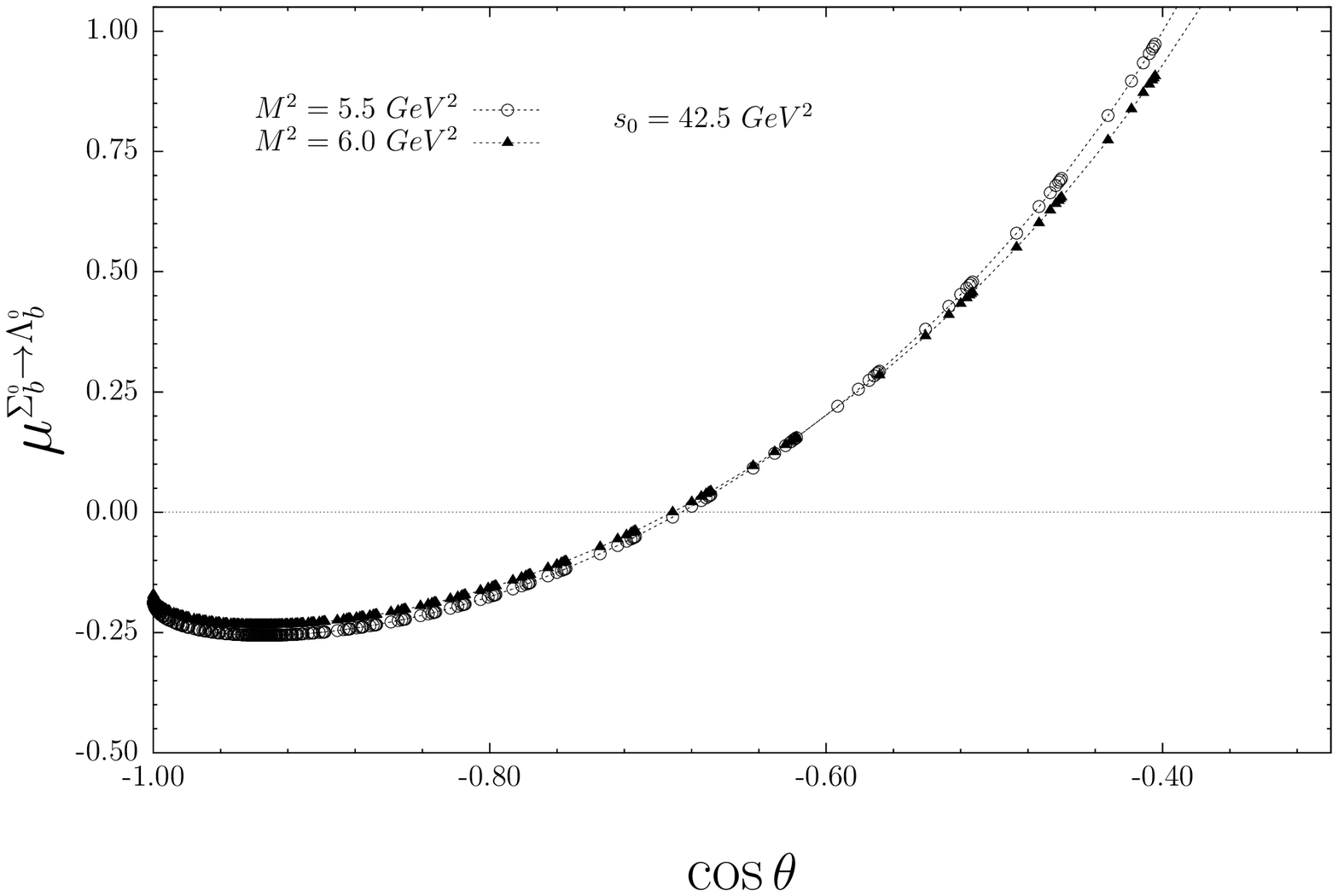}
\vskip 7.0cm
\caption{}
\end{figure}

\end{document}